\documentclass[aps,prb,showpacs,twocolumn,floats,superscriptaddress]{revtex4}
\usepackage{amssymb}
\usepackage{amsbsy}
\usepackage{amsmath}
\usepackage{epsfig}

\begin{document}

\title {Mott phases and superfluid-insulator transition of dipolar spin-three bosons
in an optical lattice: implications for $^{52}$Cr atoms}

\author{Jean-S\'ebastien Bernier}
\affiliation{Department of Physics, University of Toronto,
Toronto, Ontario, Canada M5S 1A7}
\author{K. Sengupta}
\affiliation{TCMP Division, Saha
Institute of Nuclear Physics, 1/AF Bidhannagar, Kolkata-700064, India}
\author{Yong Baek Kim}
\affiliation{Department of Physics, University of Toronto,
Toronto, Ontario, Canada M5S 1A7}
\affiliation{School of Physics, Korea Institute for Advanced Study, Seoul
130-722, Korea}
\affiliation{Department of Physics, University of California, Berkeley,
California 94720}

\date{\today}

\begin{abstract}

We study the Mott phases and superfluid-insulator transition of
spin-three bosons in an optical lattice with an anisotropic two
dimensional optical trap. We chart out the phase diagrams for Mott
states with $n=1$ and $n=2$ atoms per lattice site. It is shown that
the long-range dipolar interaction stabilizes a state where the
chains of the ferromagnetically aligned spins run along the longer
trap direction while the spin ordering is staggered between nearby
chains, leading to an antiferromagnetic ordering along the shorter
trap direction. We also obtain the mean-field phase boundary for the
superfluid-insulator transition in these systems and study the
nature of spin ordering in the superfluid state near the transition.
We show that, inside the superfluid phase and near the
superfluid-insulator phase boundary, the system undergoes a first
order antiferromagnetic-ferromagnetic spin ordering transition. We
discuss implications of our results for $^{52}$Cr atoms and suggest
possible experiments to detect different phases in such systems.

\end{abstract}


\maketitle

\section {Introduction}

Several experiments on ultracold trapped atomic gases have opened a
new window onto a plethora of intriguing phases of
quantum matter. \cite{greiner1,orzel1}
A gas of bosonic atoms in an optical or magnetic trap has been
reversibly tuned  between superfluid (SF) and insulating ground
states by varying the strength of a periodic potential produced by
standing optical waves. This transition has been explained on the
basis of the Bose-Hubbard model with on-site repulsive interactions
and hopping between nearest neighboring sites of the lattice
\cite{fisher1}. As long as the atom-atom interactions are small
compared to the hopping amplitude, the ground state remains
superfluid. In the opposite limit of a strong lattice potential, the
interaction energy dominates and the ground state is a Mott
insulator (MI) when the density is commensurate, with an integer
number of atoms localized at each lattice site.

In the above-mentioned experiments, the spins of the atoms were
neglected either due to freezing of the spin degrees of freedom in
the presence of a magnetic trap or due to the use of spinless atoms.
Later, use of optical trap for confining atoms led to realization of
spin-one and spin-two bosonic systems where the spins of the atoms
are dynamical degrees of freedom \cite{xx1,xx2}. Furthermore,
possibilities of creating two-component bosonic mixture, which is
equivalent to bosons with pseudospin half, has also been discussed
\cite{demler1}. In all of these systems, the bosonic atoms are shown
to have spin-dependent contact interaction which can lead to variety
of spin ordered Mott ground states
\cite{issacson1,demler1,zhou1,subir1,barnett,demler2}. The superfluid-insulator
transitions in these systems has also been studied
\cite{issacson1,demler2,subir1,hou}.

More recently, a Bose-Einstein condensate of $^{52}$Cr atoms has
been realized \cite{pfau1}. These Cr atoms, in contrast to their
alkali counterparts, have the electronic configuration $[Ar]\,3d^5\,4s^1$
which leads, for the isotopes with no nuclear spin, to a net spin of $3$ (in units of $\hbar$)
and to a magnetic moment of $6\mu_B$, where $\mu_B$ is the Bohr magneton
\cite{pfau2}. Consequently, these atoms experience a much stronger
long-range dipolar interaction compared to their alkali
counterparts. Besides, again in stark contrast to other bosonic alkali atoms,
in a Cr condensate, the scattering lengths in different angular momentum
channel are quite different\cite{ho1}.
Since the Cr atoms have spin 3, the relevant scattering
lengths ($a_S$) occurs in angular momentum channels $S=0,2,4,6$. All
the $a_S$ except $a_0$ are determined and have the values
$a_2=-7a_B$, $a_4=58a_B$ and $a_6=112 a_B$, where $a_B$ is the Bohr
radius \cite{ho1,pfau2}. Such different scattering lengths lead to
much stronger spin-dependent contact interaction potentials in Cr
compared to its alkali counterparts \cite{ho1}. In fact, in the
presence of an optical trap, where the spins become dynamical
degrees of freedoms, the spin-dependent contact interactions
overwhelm the long-ranged dipolar interaction energy, and lead to
various spin-nematic phases \cite{ho1,santos1}. In particular, it
has been predicted in Ref.\ \onlinecite{ho1} that biaxial nematic
phases may be realized in Cr condensates \cite{comment1}.

In this work, we study the Mott phases and superfluid-insulator
transition of spin-three bosons in an optical lattice with
an anisotropic two dimensional optical trap. The main results
reported in this paper are as follows. First, we map out the
phase diagrams for Mott states with $n=1$ and $n=2$ atoms per
lattice site. In particular, we demonstrate that both the contact
and the dipolar interactions are important for obtaining the correct
Mott ground states. It is shown that the presence of a long-range
dipolar interaction and an anisotropic two-dimensional (2D) trap
lead to a Mott ground state where
the chains of the ferromagnetically (FM) aligned
spins run along the longer trap direction while the spin ordering is
staggered between nearby chains, leading to an antiferromagnetic (AFM)
ordering along the shorter trap direction.
Second, we obtain the superfluid-insulator
transition (SIT) phase boundary of this system using mean-field
theory and discuss the spin ordering of the superfluid near and away
from the transition. We demonstrate that, in contrast to the alkali
bosonic atoms with weak spin-dependent contact interaction and
negligible dipolar interactions, the spin-three bosons with dipolar
interaction exhibit a first order AFM-FM transition for the spin ordering
in the condensate as one moves away from the SIT boundary into the
superfluid phases. Finally, we suggest future experiments
on $^{52}$Cr to test our theoretical predictions.

We start with the following Hamiltonian
\begin{eqnarray}
{\mathcal H}&=& {\mathcal H'_K}+{\mathcal H'_O}+{\mathcal H'_D}
\label{ham1} \\
 {\mathcal H'}_K&=& \int d{\bf r}~\hat{\psi}^{\dagger}_{a}({\bf r})
\left(-\frac{\hbar^2}{2M}\nabla^2 + V({\bf r}) \right)
\hat{\psi}_{a}({\bf r}) \label{hamk} \\
{\mathcal H'}_O&=& \frac{1}{2} \int d{\bf r}~d{\bf r'}
\hat{\psi}^{\dagger}_{a}({\bf r}) \hat{\psi}^{\dagger}_{a'}({\bf
r'}) u({\bf r}-{\bf r'}) \hat{\psi}_{b'}({\bf r'})
\hat{\psi}_{b}({\bf r}) \label{hamonsite} \\
 {\mathcal H'}_D &=& \frac{u_4}{2} \int
d{\bf r}~d{\bf r}' \hat{\psi}^{\dagger}_{a}({\bf r})
 \hat{\psi}^{\dagger}_{a'}({\bf r'}) \left(\frac{{\bf S}_{a b}
\cdot {\bf S}_{a' b'}}{|{\bf r}-{\bf r'}|^3}
\right.\nonumber\\
&& \left. - 3~S_{a b}^u ~S_{a' b'}^v~\frac{({\bf r}-{\bf
r'})_u~({\bf r}-{\bf r'})_v}{|{\bf r}-{\bf r'}|^5}\right)
\hat{\psi}_{b'}({\bf r'}) \hat{\psi}_{b}({\bf r}), \label{mbH}
\nonumber\\
\end{eqnarray}
where $\hat{\psi}^{\dagger}_{a}({\bf r})$ is the boson creation
operator with spin projection $a=\{-3,-2,-1,0,1,2,3\}$ at position
${\bf r}$, $V({\bf r})$ is the optical lattice potential, $S_{a
b}^u$ denotes the element of the three spin-3 rotation matrices, sum
over all repeated indices are implied, and $u({\bf r}-{\bf r'})$ is
the two-body interatomic potential given by
\begin{equation}
u({\bf r} - {\bf r'})=\delta({\bf r} - {\bf r'})(g_0 P_0+g_2 P_2+g_4
P_4+g_6 P_6), \label{ci}
\end{equation}
where $g_S=4\pi\hbar^2 a_S/M$ and $P_S=\sum_{m=-S}^S |S,m\rangle
\langle S,m|$ is the projection operator for the Bosons. Using the
identities
\begin{eqnarray}
1 &=& P_0+P_2+P_4+P_6,\nonumber\\
({\bf S} \cdot {\bf S'}) &=& -12 P_0-9 P_2-2 P_4+9 P_6, \nonumber\\
({\bf S} \cdot {\bf S'})^2 &=& 144 P_0+81 P_2+ 4 P_4+81 P_6,
\end{eqnarray}
we can rewrite the contact interaction as
\begin{equation}
u({\bf r} - {\bf r'})=\delta({\bf r} - {\bf r'})\left[u_0 +u_1 P_0
+u_2 ({\bf S} \cdot {\bf S'})+u_3 ({\bf S} \cdot {\bf S'})^2
\right],
\end{equation}
where $u_0=(-11 g_2+81 g_4 +7 g_6)/77$, $u_1=g_0+(-55 g_2+27 g_4-5
g_6)/33$, $u_2=(g_6-g_2)/18$, $u_3=(g_2/126-g_4/77+g_6/198)$. The
coefficient of the dipolar term $u_4$ is known to be small (${\it
i.e.}$ $u_4 \ll u_0,u_1,u_2,u_3$) in $^{52}$Cr atom\cite{ho1}.

For ultracold atoms in an optical lattice, it is well-known that the
energy eigenstates are Bloch wave functions and a superposition of
these Bloch states yields a set of Wannier functions that are well
localized on the individual lattice sites for deep enough lattices
\cite{greiner1,jaksch}. Further, in these systems both the kinetic and
interaction energies involved in the dynamics are small compared to
the excitation energies to higher single particle bands in the limit
of deep optical lattices necessary to realize the Mott insulating
state \cite{greiner1,jaksch}. We shall therefore expand the boson field
operators in the Wannier basis keeping only the lowest single
particle band, $\psi_a({\bf r})=\sum_i b_{ia} w({\bf r}-{\bf r}_i)$,
where $b_{ia}$ is the spin-3 boson annihilation operator at site $i$
with spin projection $a$. After a few straightforward manipulations,
we then obtain the effective single band extended Bose-Hubbard
Hamiltonian on a lattice ${\mathcal H}_B = {\mathcal H}_K +
{\mathcal H}_O + {\mathcal H}_D$, where
\begin{eqnarray}
{\mathcal H}_K &=& -t \sum_{\langle ij \rangle}\sum_{a} \left(
b_{ia}^{\dagger} b_{ja} + { \rm h.c} \right)
\label{kinetic1} \\
 {\mathcal H}_O &=& \frac{U_0}{2} \sum_{i} {\hat n}_{i} ({\hat n}_{i}-1)
+\frac{U_2}{2} \sum_{i} ({\bf S}_i^2 - 12~\hat{n}_i) \nonumber\\
&& + \frac{U_1}{2} \sum_{i} \sum_{a b} \frac{1}{7}
(-1)^{a}~b_{ia}^{\dagger}~b_{i -a}^{\dagger} ~(-1)^{b}~b_{ib}~b_{i
-b} \nonumber\\
&& +\frac{U_3}{2} \sum_{i} [ \sum_{k,l=\{x,y,z\}}
N_{i,kl}^2 -132~\hat{n}_i ] \label{interact1} \\
{\mathcal H}_D &=&  \frac{U_4}{2} \sum_{i,j}\left(\frac{{\bf S}_i
\cdot {\bf S}_j}{|{\bf r}_i-{\bf r}_j|^3} \right.
\nonumber\\
&& \left. -3~S_i^v~S_j^v \frac{({\bf r}_i-{\bf r}_j)_u~({\bf
r}_i-{\bf r}_j)_v} {|{\bf r}_i-{\bf r}_j|^5}\right). \label{dipole}
\end{eqnarray}
Here ${\hat n}_i = \sum_{a} b_{ia}^{\dagger} b_{ia}$ is the boson
density at site $i$, ${\bf S}_i = \sum_{a b} b_{ia}^{\dagger} {\bf
S}_{ab} b_{ib}$ is the spin operator and $N_{i,kl}=\sum_{a b}
b_{ia}^{\dagger} (S_k S_l)_{ab} b_{ib}$ is the `nematic' operator.
The hopping coefficient $t$ is given by $t=\int d{\bf r}~w^{*}({\bf
r}-{\bf r}_i) \left(-\frac{\hbar^2}{2M}\nabla^2 + V({\bf r})
\right)~w({\bf r}-{\bf r}_j)$ and the $U$'s are given by $U_i= u_i
\int d{\bf r} |w({\bf r})|^4$, for $i=\{0,1,2,3\}$, and 
$U_4 = u_4 \int d{\bf r} |w({\bf r})|^2 ~\int d{\bf r'} |w({\bf r'})|^2$.

The Hamiltonian ${\mathcal H}_B$ is the starting point of our study
in the subsequent sections. In what follows, we shall obtain the
ground states of $ {\mathcal H}_B -\mu \sum_i {\hat n}_i$
as a function of
$U_1/U_0$, $U_2/U_0$, $U_3/U_0$, $\mu/U_0$, and $t/U_0$, where
$\mu$ denotes the chemical potential for a fixed weak dipolar
interaction $U_4/U_0 \ll 1,U_{i=\{1, 2, 3\}}/U_0$. Note that
since the value of $a_0$ and hence $U_1$ is not experimentally
determined, the position of $^{52}$Cr in this generalized phase diagram is
not known.
%
%
However, we expect that some of the general features
obtained from this phase diagram will apply to $^{52}$Cr. Also,
we would like to point out that in principle we expect the chemical
potential $\mu$ to be space-dependent due to the presence of the
trapping potential. In this work, we have ignored this effect of the
optical trap which is a standard approximation in the literature for
large traps \cite{florian1}.

The rest of the paper is organized as follows. In Sec.
\ref{secmott}, we study various possible Mott phases of the ${\mathcal H}_B$.
This is followed, in Sec.\ \ref{secsit}, by a mean-field study of
the SIT in this model. Then in Sec.\ \ref{secexp},
we suggest experiments to test our predictions. Finally, we conclude
in Sec.\ \ref{seccon}.

\section{Mott phases}
\label{secmott}

In this section, we obtain various possible Mott phases when
there are integer $n$ particles per site. In the Mott limit, the
kinetic energy of atoms are negligible and we shall take $t=0$ in
the rest of this section. We shall concern ourselves only with the
case of $n=1$ and $n=2$ atoms per site in this work. The
calculations can be generalized, in principle, to higher $n$ values.
The phase diagram for the on-site terms is sketched in Sec.\
\ref{submotta} while the effect of the dipolar interaction is
presented in Sec.\ \ref{disec}.

\subsection{On-site Interaction}
\label{submotta}

In order to obtain the Mott phases, we use the
fact that the dipolar interaction term ${\mathcal H}_D$ (Eq.\
\ref{dipole}) is small compared to the on-site interaction terms in
${\mathcal H}_O$ (Eq.\ \ref{interact1}). Thus, we first neglect the
dipole term ${\mathcal H}_D$ and find out the ground-state of
${\mathcal H}_O$ by  minimizing the variational energy
\begin{eqnarray}
E_v (n) &=& \left<\Psi_v(n)\right| {\mathcal
H}_O\left|\Psi_v(n)\right> \label{evar},
\end{eqnarray}
where the on-site variational wavefunction $|\Psi_v(n)\rangle$ for
$n$ bosons per lattice site is given by
\begin{eqnarray}
|\Psi_v (n)\rangle &=& \prod_i \sum_{S=S_{\rm total}}
\sum_{m=-S}^{S} c_{i,(n,S,m)} |n; S,m\rangle_i,
\end{eqnarray}
where $S_{\rm total}$ denotes all the possible total spin that can
be obtained by adding spins of $n$ spin-three bosons and $m$ is the
azimuthal quantum number corresponding to $S$. For example for
$n=2$, the possible values of $S_{\rm total}=0,2,4,6$.

\begin{figure}
\vspace{0cm} \rotatebox{0}{
\includegraphics[width=6cm]{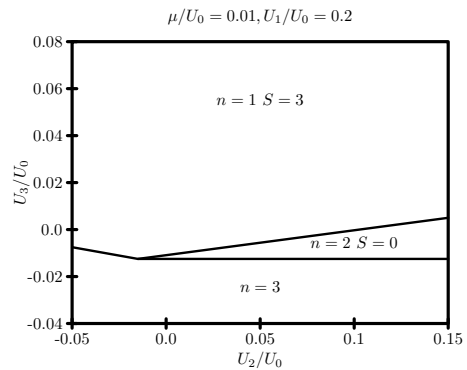}}
\caption{Mott phase diagram obtained by minimization of $E_v(n)$
for $n=1$ and $n=2$ as a function of $U_2/U_0$ and $U_3/U_0$ for
$\mu=0.01U_0$ and $U_1/U_0=0.2$. Note that the azimuthal degeneracy
is not lifted and hence the ground states for non-zero $S$ have a
$2S+1$ fold degeneracy.} \label{fig1}
\end{figure}

\begin{figure}[t]
\vspace{0cm} \rotatebox{0}{
\includegraphics[width=6cm]{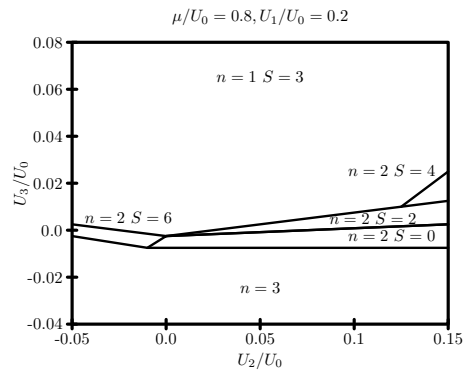}}
\caption{Same as in Fig.\ \ref{fig1} for $\mu=0.8U_0$ and
$U_1/U_0=0.2$} \label{fig2}
\end{figure}

\begin{figure}
\vspace{0cm} \rotatebox{0}{
\includegraphics[width=6cm]{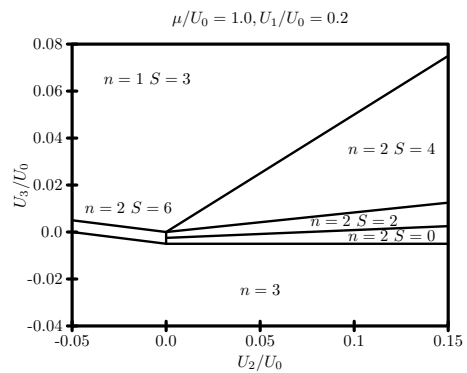}}
\caption{Same as in Fig.\ \ref{fig1} for $\mu=U_0$ and
$U_1/U_0=0.2$} \label{fig3}
\end{figure}

\begin{figure}[t]
\vspace{0cm} \rotatebox{0}{
\includegraphics[width=6cm]{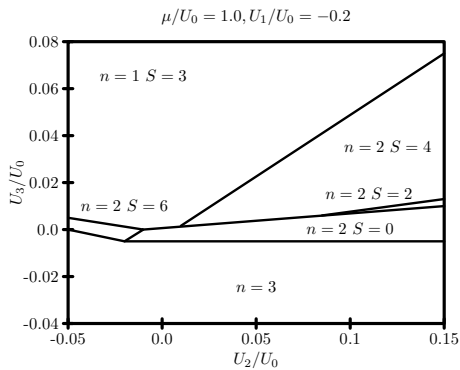}}
\caption{Same as in Fig.\ \ref{fig1} for $\mu=U_0$ and
$U_1/U_0=-0.2$} \label{fig4}
\end{figure}

Details of the calculation of $E_v(n)$ are sketched in Appendix
\ref{appa}. Here we present the phase diagram obtained by numerical
minimization of $E_v(n)$ for $n=1$ and $n=2$ as a function of
$U_3/U_0$ and $U_2/U_0$ for several representative values of
$\mu/U_0$ and for $U_1/U_0=\pm 0.2$. These values of $\mu$ and $U_1$
are chosen for clarity; we have checked that there are no other
phases that occur for other values of $\mu$ and $U_1$. In order to
find the lower boundary in $U_3/U_0$ for the $n=2$ states, we also
minimize $E_v(n=3)$. However, since we are mostly concerned with the
$n=1$ and $n=2$ states and also for clarity, we have not indicated
the different $S$ regions of $n=3$ in the phase diagrams shown in
Figs.\ \ref{fig1}, \ref{fig2}, \ref{fig3}, and \ref{fig4}. From our
calculations, we find that ${\mathcal H}_O$ do not lift the
azimuthal degeneracy of the atoms and the ground states obtained
from this on-site variational wavefunction are degenerate. The only
exception to this occurs for the singlet ground state ($S=0$) that
exists in a narrow region of the phase diagram. For all other
phases, the on-site interactions determine only the $S$ value of the
ground state, thus leaving a $2S+1$ degeneracy corresponding to
different possible $m$ values. This is where the dipole interaction
comes into play. The presence of the dipolar interaction fixes the
direction of the spins and gives us a unique ground state. Notice
that the weakness of the dipolar interactions compared to the
on-site term ${\mathcal H}_O$ ensures that the former would not
change the value of $S$ in the ground state.

\subsection{Dipolar Interaction}
\label{disec}

To investigate the effect of the dipolar interaction, we now restrict
ourselves to the lowest $S$ manifold which has been determined by
minimizing $E_v(n)$. However, even within this manifold, we need to
construct a variational wavefunction to obtain the ground state of
${\mathcal H}_D$. Since the dipole interaction is long-ranged, such a
trial wavefunction need not be a simple on-site product
wavefunction. To guess the form of this wavefunction, we first look
at the dipolar interaction for classical spins in an optical lattice
and find out the ground state for such spins. Next, we choose a
variational wavefunction which is compatible with the symmetry of
this classical ground state.

In the classical limit, the spins can be parameterized as ${\bf
S}_i=S(\sin\theta_i\cos\phi_i,\sin\theta_i\sin\phi_i,\cos\theta_i)$.
Also, due to the form of the dipolar interaction (Eq.\ \ref{dipole})
and the two-dimensional nature of our system, it is easy to see that the ground
state, in the classical limit, shall have all spins aligned in the
XY plane. Thus in the ground state, each classical spin can be represented
by a single angle $\phi$. This simplifies the computational procedure
enormously and enables us to study the ground states of classical
spins for different trap configurations with both open and periodic
boundary conditions.

\begin{figure}
\vspace{0cm} \rotatebox{90}{
\includegraphics[width=6cm]{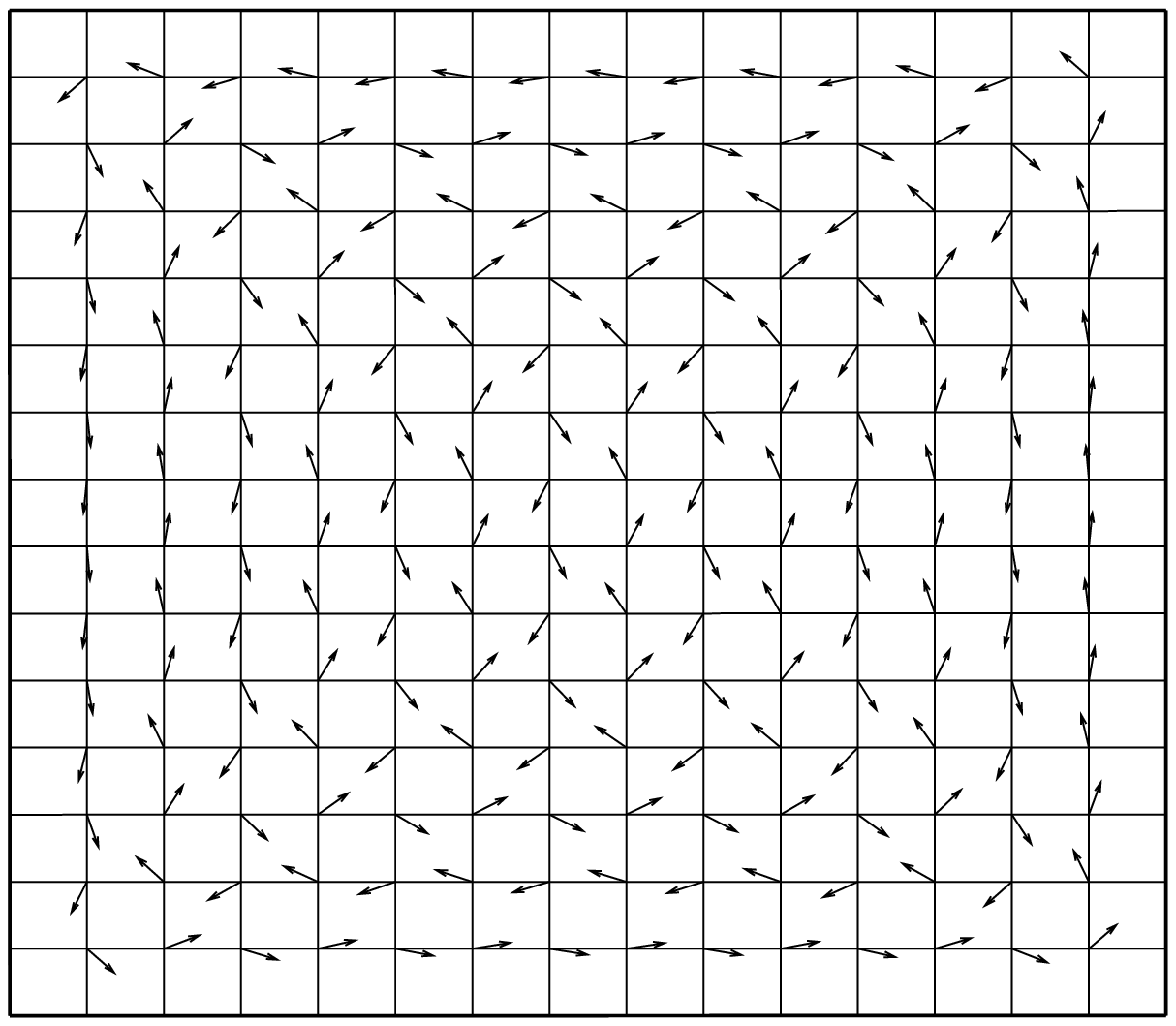}}
\caption{$14\times 14$ lattice with the open boundary condition. 
Notice that the spins in the bulk region are following a ``swirling''
pattern that can be described by four interpenetrating sublattices.}
\label{fig5}
\end{figure}

\begin{figure}
\vspace{0cm} \rotatebox{0}{
\includegraphics[width=4cm]{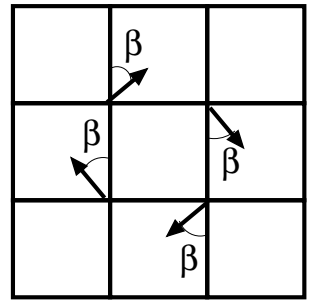}}
\caption{Typical spin pattern for classical spins with dipolar interactions
in the square lattice with the periodic boundary condition. 
The ground state has infinite degeneracy corresponding 
to different values of $\beta$.} 
\label{fig5A}
\end{figure}

\begin{figure}[t]
\vspace{0cm} \rotatebox{90}{
\includegraphics[width=4.5cm]{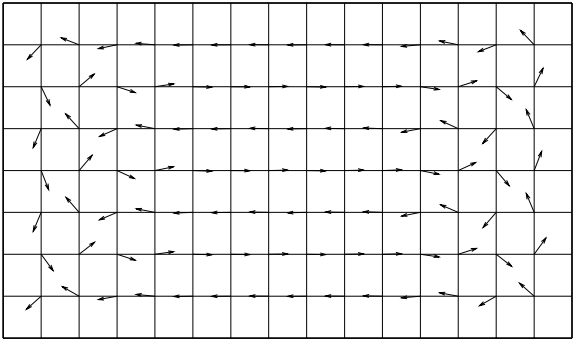}}
\caption{$14\times 7$ lattice with the open boundary condition. Notice
that the spins in the bulk region are aligned along the longer trap
direction. This pattern can be described via four interpenetrating
sublattices.} \label{fig6}
\end{figure}

To find the minimum energy configuration for the classical spins, we
minimize ${\mathcal H}_D(\phi_i)$ for both square and rectangular
trap geometries. For square traps with open boundary conditions or
rectangular traps with periodic boundary conditions in both $x$ and
$y$ directions, the bulk region exhibits a swirling pattern
described by four interpenetrating sublattices (see Fig.\ref{fig5})
with the spins making an angle $\beta$ with the lattice at each site
(see Fig.\ \ref{fig5A}). Moreover, simulations with periodic
boundary conditions in both trap directions show that the system
energy is degenerate for all values of $\beta$. Consequently, for
square traps, dipolar interactions are not sufficient to lift the
spin degeneracy. However, this situation is difficult to
realize in realistic finite size systems with optical traps where a
rectangular trap with open boundary condition is a more realistic
choice. For the rest of this work, we shall therefore concentrate on
such geometries. In this case, the spins of the bulk region tend to
align with the longest trap direction as shown in
Fig.\ref{fig6}. Consequently, in the classical sense, the spin
degeneracy is completely lifted by dipolar interactions in
rectangular geometries. The ground state for the classical spins for
rectangular trap with open boundary condition thus have an
antiferromagnetic(AFM) order along the shorter trap direction. Note
that such an order corresponds to an interpenetrating
four-sublattice structure shown in Fig.\ \ref{fig5A} with $\beta=\pm
\pi/2$ when $x$ is taken as the longest trap direction.

The lesson we learn from the study of the dipolar
Hamiltonian in the classical limit is that for rectangular traps
with open boundary conditions, the ground state has four
interpenetrating sublattices as seen in Fig.\ \ref{fig6}. Therefore,
a reasonable minimal trial wavefunction for finding the ground state
of ${\mathcal H}_D$ in the quantum limit must at least have a
four-sublattice structure. With this observation, we choose the
following variational wavefunction for $n=1$
\begin{eqnarray}
\left|\Psi_{\rm dipole}(n=1)\right>&=& \prod_{\Lambda=\{A,B,C,D\}}
~\prod_{i \in \Lambda}~\times \\ \nonumber && \sum_{m=-3}^3
c_{i,(n=1,S=3,m)} |1;3,m\rangle_i,
\end{eqnarray}
where $\{A, B, C, D\}$ are the four sublattices. Similarly for
$n=2$, one gets
\begin{eqnarray}
\left|\Psi_{\rm dipole}(n=2)\right>&=& \prod_{\Lambda=\{A,B,C,D\}}
~\prod_{i \in \Lambda}~\times \\ \nonumber && \sum_{S=\{0,2,4,6\}}
\sum_{m=-S}^S c_{i,(n=2,S,m)} |2;S,m\rangle_i.
\end{eqnarray}
We then minimize $E_v^{\rm dipole}(n) = \left< \Psi_{\rm
dipole}(n)\right|{\mathcal H}_D \left|\Psi_{\rm dipole}(n)\right>$
to obtain the ground state for the system. We have carried out this
numerical minimization for rectangular lattice sizes up to $16\times
8$ with periodic boundary conditions in one of the two directions.
For all values of $S\ne 0$ and for all lattice sizes up to $16\times
8$, we find again that the spins are aligned parallel to the longest
trap direction and have the same four interpenetrating sublattice
structure as shown in Fig.\ \ref{fig6}. Since this ground state
turns out to be generic for all $S\ne 0$, we expect to find such a
ground state for $^{52}$Cr atoms irrespective of its position in the
generalized phase diagram for $n=1$. For $n=2$, the ground state
will either be the four interpenetrating sublattice structure
mentioned previously (if $S \ne 0$) or a singlet ($S$=0).

\section{Superfluid-Insulator transition}
\label{secsit}

\subsection{Phase Boundary}
\label{pb}

In this section, we obtain the superfluid-insulator phase boundary
starting from the $n=1$ Mott phase within a mean-field theory.
To allow density fluctuations in the superfluid state,
we construct a variational wavefunction that is a
superposition of $n=0$, $n=1$ and $n=2$ Mott states. Note that the
variational wavefunction here needs to have the four sublattice
structure mentioned in Sec.\ \ref{disec} to correctly capture the
$n=1$ Mott state. Such a wavefunction is given by
\begin{eqnarray}
|\Psi\rangle_{0}&=& \prod_{\Lambda=\{A,B,C,D\}} \prod_{i \in \Lambda} \times \\ \nonumber
&& (c_{i,(n=0)} |n=0\rangle_i + |n=1\rangle_i + |n=2\rangle_i), \label{var1}
\end{eqnarray}
where
\begin{eqnarray}
|n=1\rangle_i &=& \sum_{m=-3}^{3} c_{i,(n=1,S=3,m)} b_{i,m}^{\dagger}
|0\rangle \label{n1wave}
\end{eqnarray}
and
\begin{eqnarray}
|n=2\rangle_i &=& \sum_{S=\{0,2,4,6\}} \sum_{m=-S}^{S} c_{i,(n=2,S,m)}
|2;S,m\rangle_i \nonumber \\
&=& \sum_{S=\{0,2,4,6\}} \sum_{m=-S}^{S} c_{i,(n=2,S,m)} \nonumber \\
&& \times \sum_{ab} \xi_{(S,a+b=m)}~b_{i,a}^\dagger b_{i,b}^\dagger
|0\rangle,
\label{n2wave}
\end{eqnarray}
where $\xi_{(S,a+b=m)}$ are Clebsch-Gordan coefficients. The
normalization of $|\Psi\rangle_{0}$ leads to the condition
\begin{eqnarray}
1 &=& |c_{i,(n=0)}|^2+\sum_{m=-3}^{3}|c_{i,(n=1,S=3,m)}|^2 \nonumber\\
&& +\sum_{S=\{0,2,4,6\}}\sum_{m=-S}^{S}|c_{i,(n=2,S,m)}|^2.
\end{eqnarray}

Using this wavefunction, the expectation value of the kinetic term
is given by
\begin{eqnarray}
E_{K}&=& _0\langle \Psi|{\mathcal H}_K| \Psi \rangle_0
=-t~\sum_{\langle i,j \rangle,a} 2~\Re(\Delta_{ia}^* \Delta_{ja}),
\label{ke1}
\end{eqnarray}
where the superfluid order parameter $\Delta_{ia}^* =~_0\langle
\Psi| b_{ia}^{\dagger}| \Psi \rangle_0 $ can be expressed in terms
of coefficients of the variational wavefunction $|\Psi\rangle_{0}$
as
\begin{eqnarray}
\Delta_{ia}^* &=& c_{i,(n=1,S=3,a)}^* c_{i,(n=0)} \nonumber \\
&& + \sum_{S=\{0,2,4,6\}} \sum_{m'=-S}^S \sum_{bc} \sum_{m=-3}^3
c_{i,(n=2,S,m')}^* \nonumber\\
&& \times c_{i,(n=1,S=3,m)} \xi_{(S,b+c=m')} (\delta_{ba} \delta_{cm}
+ \delta_{bm} \delta_{ca}). \label{ord1} \nonumber\\
\end{eqnarray}

Since the interaction terms ${\mathcal H}_O$ and ${\mathcal H}_D$
(Eqs. \ref{interact1} and \ref{dipole}) do not mix states with
different $n$, the total variational energy can be written as
\begin{eqnarray}
E_{\rm var} &=& E_{Mott}(n=1) + E_{K} + \delta E_h + \delta E_p, \nonumber\\
\label{varen1}
\end{eqnarray}
where $\delta E_p$ and $\delta E_h$ which are the energy costs of
adding a particle and hole respectively to the Mott state, are given
by
\begin{eqnarray}
\delta E_p &=& \frac{1}{N} \Big[\langle \Psi_{\rm dipole}(n=2)| \nonumber \\
&& \times \Big( {\mathcal H}_O + {\mathcal H}_D - \mu \sum_i {\hat n}_i \Big)
|\Psi_{\rm dipole}(n=2)\rangle  \nonumber \\
&& - \langle \Psi_{\rm dipole}(n=1)| \nonumber \\
&& \times \Big( {\mathcal H}_O + {\mathcal H}_D - \mu \sum_i {\hat n}_i \Big)
|\Psi_{\rm dipole}(n=1)\rangle \Big]  \nonumber \\
\label{dep} \\
\delta E_h &=& - \frac{1}{N} 
\langle \Psi_{\rm dipole}(n=1)| \nonumber \\
&& \times \Big( {\mathcal H}_O + {\mathcal H}_D - \mu \sum_i {\hat n}_i \Big)
|\Psi_{\rm dipole}(n=1)\rangle.
\nonumber \\
\label{deh}
\end{eqnarray}
The mean-field ground state of the system can now be obtained, as a
function of $t/U_0$ and $\mu/U_0$ for representative values of
$U_1/U_0$, $U_2/U_0$, $U_3/U_0$ and $U_4/U_0$, by numerical
minimization of $E_{\rm var}$. Note that as $t$ exceeds a critical
value $t_c$, the energy gain from $E_K$ (Eq.\ \ref{ke1}) exceeds the
energy cost of adding a particle ($\delta E_p$)  or a hole ($\delta
E_h$) to the Mott state. At this point, the ground state of the
system occurs with non-zero superfluid order parameter $\Delta_a = \sum_i \Delta_{ia}$.

We obtain $t_c$ by minimizing $E_{\rm var}$ (Eq.\ \ref{varen1}). In
doing so, to reduce computational time, we only keep the two lowest
manifolds of $S$ states in the Mott phase corresponding to
$|n=2\rangle_i$ (Eq.\ \ref{n2wave}). The justification of this
procedure comes from the following. At the superfluid-insulator
phase boundary, the hopping coefficient $t$ reaches a critical value
for which it becomes energetically favorable to add (or remove) a
particle to the $n=1$ Mott state. When such a particle is added, the
resultant state has $n=2$ and can, in principle, have superposition
of $S=0,2,4,6$ states due to the presence of the dipole term.
However, when the dipole interaction is weak compared to the on-site
interaction, such superposition becomes energetically costly. Thus
the resultant lowest-energy $n=2$ state is almost entirely localized
within the two lowest $S$ manifolds for $t \simeq t_c$. We have
numerically checked for a few cases keeping all the $S$ states that
this is indeed the case.

\begin{figure}[t]
\vspace{0cm} \rotatebox{270}{
\includegraphics[width=6cm]{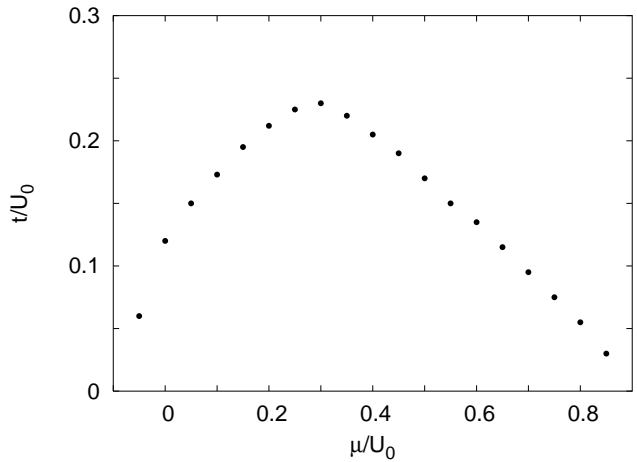}}
\caption{The mean-field phase boundary for the superfluid-insulator
transition with the $n=1$ Mott state in an optical lattice. Here $U_2=0.05 U_0$,
$U_3=0.02 U_0$ and $U_4=0.0125U_0$. Notice the relatively higher
values of $t_c/U_0$ compared to the usual cases, which is a consequence of 
spin order in the Mott phase (see text).} \label{fig7}
\end{figure}

The mean-field superfluid-insulator phase boundary, so obtained, is
shown in Fig.\ \ref{fig7} for representative parameter values
$U_1=0$, $U_2=0.05 U_0$, $U_3=0.02 U_0$ and $U_4=0.0125 U_0$. Here
the two lowest $n=2$ states corresponds to $S=2$ and $S=4$ manifolds
and we have kept these states for our minimization calculation. We
note that compared to the usual Bose-Hubbard model \cite{fisher1},
the transition occurs for a relatively higher $t_c/U$. This
phenomenon can be understood to be the consequence of the
spin-ordering of the Mott state which originates from the dipole
interaction. For sake of clarity, we shall consider the situation
where the Mott state is destabilized by addition of holes. The other
case, where the Mott state is destabilized by addition of particles,
has a similar explanation.

Consider the Mott state of the system with the spin order as shown
in Fig.\ \ref{fig6}. It turns out that the coefficients $c_{i}$ corresponding
to spins in nearby chains have opposite signs because the
spin directions are staggered from one chain to the other.
Since
$\Delta_{i,a}^* \sim c_{i,(n=1,S=3,a)}^* c_{i,(n=0)}$, the sign of
$\Delta_i$ varies between different sites.
Consequently, the sign of $\Delta_i^{\ast} \Delta_j$ in Eq.\ \ref{ke1}
can vary so that the kinetic energy is `frustrated'.
This leads to a reduction of the kinetic energy and hence one
needs a higher $t$ value to destabilize the Mott phase. We note that
this phenomenon is generic and is a consequence of dipolar interaction
between the spins, which stabilizes the above-mentioned four-sublattice order.

\subsection{AFM-FM transition}

In this section, we study the spin-ordering of the superfluid phase
near the transition. The strategy for such a study is again
minimization of $E_{\rm var}$ followed by computation of $\langle S_x
\rangle$ and $\langle S_y \rangle$ using the variational ground
state wave function. However, even before such a computation is
carried out, one can guess the result from a simple physical
picture, which we now present.

\begin{figure}[t]
\vspace{0cm} \rotatebox{270}{
\includegraphics[width=6cm]{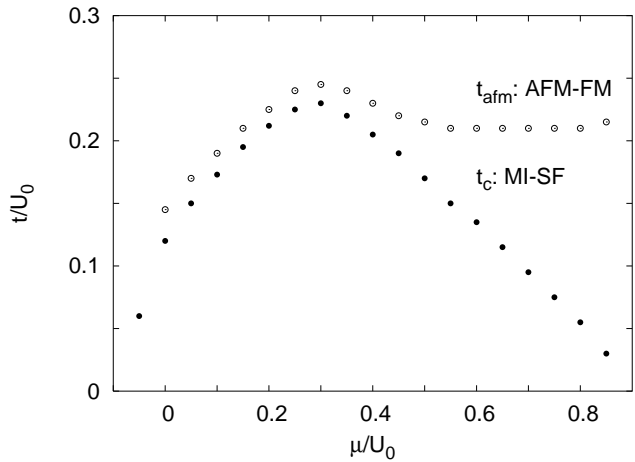}}
\caption{The AFM-FM and the MI-SF transition boundaries for the 
$n=1$ Mott state in an optical lattice as a function 
of $\mu/U_0$. Here $U_2=0.05 U_0$, $U_3=0.02 U_0$ and
$U_4=0.0125U_0$. Notice that the AFM-FM transition follows closely
the MI-SF transition for values of $\mu/U_0$ smaller than
$(\mu/U_0)_{t_{c}^{\rm max}}$ and departs from the MI-SF transition
boundary for larger values of $\mu/U_0$.} \label{afm-fm}
\end{figure}

We have already noted in Sec.\ \ref{pb} that the AFM spin-order
along the shorter trap direction, which minimizes the dipole energy
in the Mott state, is in competition with maximization of kinetic
energy near the superfluid-insulator transition. Thus as we go
deeper in the superfluid phase, we expect that there would be a
transition from a state with such a spin-order to a FM state where
all the spins point towards $x$ (the longer trap direction).
Such a transition from the translational-symmetry broken AFM state
to a uniform FM state is expected to be first order. The exact
location of this transition depends on the relative strength of
$t$, $\mu$ and $U_4$. As shown in Fig.\ \ref{afm-fm}, the AFM-FM
transition follows closely the MI-SF transition for values of
$\mu/U_0$ smaller than $(\mu/U_0)_{t_{c}^{\rm max}}$ and departs
from the MI-SF transition boundary for larger values of $\mu/U_0$.
This departure from the MI-SF boundary can be explained by noticing
that the Mott state is destabilized primarily by addition of holes
(particles) for $\mu/U_0$ close to $0(1)$. For intermediate values of
$\mu$, the destabilization of the Mott phase occurs, within
mean-field theory, with finite amplitudes for addition of both holes
and particles. Consequently, for a fixed $\mu/U_0$ close to $0$,
as we increase $t/U_0$, the average number of bosons per site
decreases and $\langle S_x \rangle$ decreases accordingly. 
Therefore, the effective dipole interaction becomes weaker and $t_{\rm afm}$, the
value of $t$ for which the kinetic energy gain wins
over the optimization of the dipolar energy, stays close to $t_c$. 
However, for a fixed $\mu/U_0$ close to $1$, the average number of
bosons per site increases as $t$ is increased beyond $t_c$
and $\langle S_x \rangle$ also slowly increases. Thus, for $\mu > \mu_{t_c^{\rm max}}$, 
the dipole interaction is robust to an increase in $t/U_0$ so that
$t_{\rm afm}$ becomes larger and deviates significantly from 
the Mott-superfluid phase boundary.

\begin{figure}[t]
\vspace{0cm} \rotatebox{270}{
\includegraphics[width=6cm]{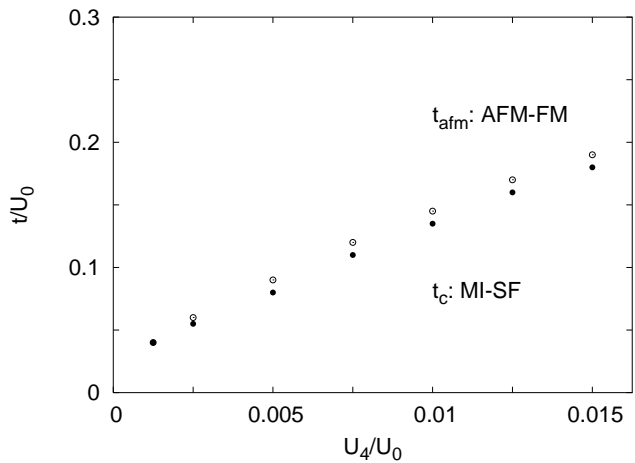}}
\caption{The AFM-FM and the MI-SF transition boundaries for the $n=1$ Mott state in an
optical lattice as a function of $U_4/U_0$ for $U_3 = 0.02 U_0$, $U_2 = 4 U_4$ 
and $\mu = 0.05 U_0$. Notice that for $U_4 = 0.00125 U_0$, the AFM-FM transition occurs 
simultaneously with the MI-SF transition.}
\label{variU4}
\end{figure}

We have also computed the variation of $t_{\rm afm}$ with $U_4$, as
shown in Fig.\ \ref{variU4} for representative values of parameters
$U_3 = 0.02 U_0$, $U_2 = 4 U_4$ and $\mu = 0.05 U_0$. We find
that $t_{\rm afm}$ approaches $t_c$ for small values of $U_4/U_0$.
The AFM-FM transition occurs simultaneously
with the MI-SF transition for $U_4 \le 0.00125 U_0$,
leading to a first order MI-SF transition in this regime.
Such dependence of $t_{\rm afm}$ on $U_4$ can
be understood by noting that a reduction of $U_4$ 
directly decreases the dipolar energy. 
As a result, when $U_4$ is reduced, the AFM spin ordering becomes less robust to 
increases in $t$ because it becomes easier to take advantage of
the kinetic energy gain at the expense of the dipolar energy.
Thus $t_{\rm afm}$ decreases, and ultimately 
merges to $t_c$ for $U_4/U_0 \le 0.00125$.

\begin{figure}[t]
\vspace{0cm} \rotatebox{270}{
\includegraphics[width=6cm]{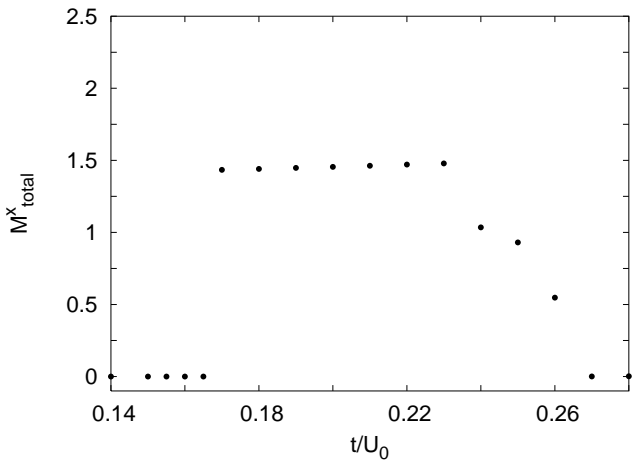}}
\caption{Total magnetization in the $x$ direction 
($M_{\rm total}^x = \langle \sum_i S_i^x \rangle/N$, where the
sum is over all lattice sites) for the Mott and superfluid 
phases as a function of $t/U_0$.
Here $\mu=0.05U_0$, $U_2=0.05 U_0$, $U_3=0.02 U_0$, 
and $U_4=0.0125U_0$. Note that the AFM-FM transition occurs for $t_{\rm afm}
\simeq 0.17U_0 > t_c \simeq 0.16U_0$ and, as we increase $t$, the FM 
order gives away to a paramagnetic superfluid for $t>0.27U_0$.} \label{fig8}
\end{figure}

\begin{figure}[t]
\vspace{0cm} \rotatebox{0}{
\includegraphics[width=9 cm]{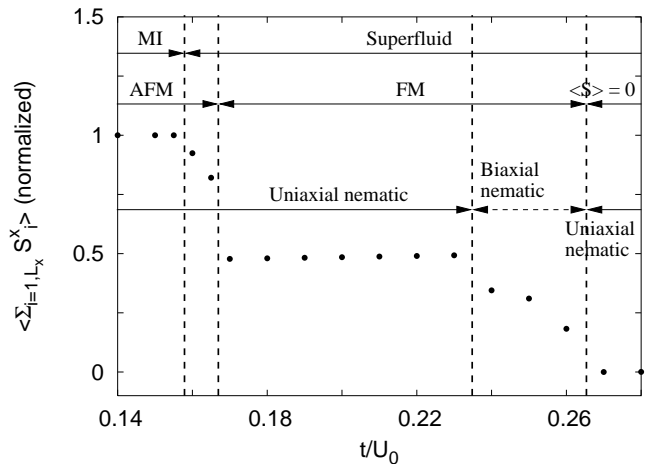}}
\caption{AFM, FM, superfluid and nematic phases as a function of
$t/U_0$ for $\mu=0.05U_0$, $U_2=0.05 U_0$, $U_3=0.02 U_0$ and
$U_4=0.0125U_0$. Here we have plotted 
$ \langle \sum_{i=1,L_x} S_{i}^x \rangle/(3 N_x)$, where the sum is 
over the $N_x$ number of sites in a given chain along the $x$ direction 
(the longer trap direction). It turns out that its magnitude in the superfluid 
phase varies with $\mu$ and the $U$'s.} \label{fig9}
\end{figure}

Finally, we investigate the fate of the spin-order as we go deeper
in the SF region. We note at the outset that since the approximation
involved in our minimization scheme loses its accuracy as we go
deeper into the superfluid phase, we do not expect our analysis to
be quantitatively accurate in this regime, but only expect some
qualitative features to be captured. To this end, we plot  $\langle
S_x \rangle$ as a function of $t/U_0$ with $\mu/U_0 = 0.05$ (with
all other parameters being the same as in Fig.\ \ref{fig7}) as shown in
Fig.\ \ref{fig8}. Note that here $t_c \simeq 0.16U_0$ and $t_{\rm afm} \simeq 0.17
U_0$, so that $t_c$ and $t_{\rm afm}$ are quite close to each other. As
we further increase $t$, we find that for $t/U_0 \ge 0.27$, we have
a transition to the paramagnetic phase where $\langle S_x \rangle =
0$. We have also computed the expectation value of the nematic order
parameters in the FM state. The results of these computation are
summarized in Fig.\ \ref{fig9}. We find that for $t\le 0.23 U_0$ and
$t\ge 0.27 U_0$, we have a uniaxial nematic phase. For
$0.23<t/U_0<0.27$, in the region over which $\langle S_x \rangle$
decrease from its maximum value to $0$, we find a biaxial nematic
phase, similar to one found in Ref.\ \onlinecite{ho1}. However, we
note that at the point where the biaxial nematicity sets in, we are
already deep inside the SF phase and our approximations of retaining
the two lowest manifold of states may be inaccurate.

\section{Experiments}
\label{secexp}

The traditional way of examining the existence of superfluidity in
trapped boson systems is to switch off the trap, let the cloud of
atoms expand freely and image the expanding cloud. The momentum
distribution of the atoms inside the trap can then be inferred by
looking at their position, or equivalently density, distribution in
the expanded cloud. Since the momentum distribution function of the
atoms is characterized by the presence/absence of coherence peaks in
the superfluid/Mott insulating states, such a measurement serves as
a qualitative probe of the state of the atoms inside the trap
\cite{greiner1}. In our proposed setup, however, such a simple
expansion alone, which is not sensitive to the different
spin states of atoms, will not be able to distinguish between all
the different phases. To achieve this distinction, we need to use
the well-known technique of passing the expanding cloud through a
Stern Gerlach magnet with the magnetic field along $y$ (shorter trap
direction) \cite{xx1,xx2}. Such a Stern-Gerlach magnet will split
the expanding cloud into two clouds of equal proportion if the atoms
in the trap were in the AFM phase. If the atoms were in the FM
phase, no such splitting will occur. Thus, a momentum-distribution
measurement, along with the Stern Gerlach measurement described
above can distinguish between the Mott-AFM (delocalized momentum
distribution and two clouds), superfluid-AFM (localized momentum
distribution and two clouds) and superfluid-FM phases (localized
momentum distribution and single cloud). However, we note that such
measurements do not uniquely determine the details of the spin
configuration in the AFM state. The signature of the details of the spin
configuration can be obtained experimentally by
measuring the spatial noise correlations of the expanding
clouds \cite{altman1}. Finally, the uniaxial and biaxial nematic
orders can also be detected using an imaging technique suggested
in Ref.\ \onlinecite{carusotto}.

\section{Conclusion}
\label{seccon}

In conclusion, we have studied the Mott insulating phases and
superfluid-insulator phase transitions of $S=3$ bosons in an optical
lattice. In particular, we have considered dipolar interaction
between the bosons and the effect of rectangular optical trap.
We have investigated possible Mott phases of such systems for
integer $n=1$ and $n=2$ atoms per lattice site and shown that the
presence of dipolar interaction leads to a state with ferromagnetic
chains along the longer trap direction which are aligned antiparallel to
each other, thus leading to an antiferromagnetic order along the
shorter trap direction. We have also presented a mean-field phase
diagram for the superfluid-insulator transition in these systems and
have shown that there is a generic first order AFM-FM transition in
the superfluid region close to the superfluid-insulator phase
boundary. We have also suggested realistic experiments
on $^{52}$Cr to verify our predictions.

\acknowledgments

This work was supported by NSERC, the Canada Research Chair Program,
the Canadian Institute for Advanced Research, KRF-2005-070-C00044
(J.S.B. and Y.B.K.), Le Fonds Qu\'eb\'ecois de la Recherche sur la
Nature et les Technologies (J.S.B.) and the Miller Institute of Basic
Research in Science at UC Berkeley via the Visiting Miller
Professorship (Y.B.K.).

\appendix
\section{Computation of the variational energy for the Mott phases}
\label{appa}

We present in this section the expression of the variational
energy $E_v(n)$ (Eq.\ \ref{evar}) in greater detail. Following Eq.\ \ref{interact1},
the on-site energy is given by
\begin{eqnarray}
E_v (n) &=& \left<\Psi_v(n)\right| \{
~\frac{U_0}{2} \sum_{i} {\hat n}_{i} ({\hat n}_{i}-1) \nonumber \\
&+& \frac{U_2}{2} \sum_{i} ({\bf S}_i^2 - 12~\hat{n}_i) \nonumber \\
&+& \frac{U_1}{2} \sum_{i} \sum_{a b} \frac{1}{7}
(-1)^{a}~b_{ia}^{\dagger}~b_{i -a}^{\dagger} ~(-1)^{b}~b_{ib}~b_{i-b} \nonumber \\
&+& \frac{U_3}{2} \sum_{i} \ [ \sum_{k,l=\{x,y,z\}} N_{i,kl}^2 -132~\hat{n}_i ] \nonumber \\
&-& \mu~{\hat n}_{i}~ \} \left|\Psi_v(n)\right>. \nonumber \\
\end{eqnarray}

To compute the on-site energy in the case of $n=1$, we use the
variational wavefunction
\begin{equation}
|\Psi_v(1)\rangle = \prod_i \sum_{m=-3}^3 c_{i,(n=1,S=3,m)} b^\dagger_{i,m} |0\rangle,
\label{wf1}
\end{equation}
while for $n=2$, we use
\begin{eqnarray}
|\Psi_v(2)\rangle &=& \prod_i \sum_{S=\{0,2,4,6\}} \sum_{m=-S}^S c_{i,(n=2,S,m)} \nonumber \\
&& \times \sum_{ab}~\xi_{(S,a+b=m)}~b^\dagger_{i,a}~b^\dagger_{i,b}|0\rangle, \nonumber \\
\label{wf2}
\end{eqnarray}
where $\xi_{(S,a+b=m)}$ are Clebsh-Gordan coefficients. Using these
wavefunctions, we find the energies for $n=1$ and $n=2$. For $n=1$,
the on-site energy is given by
\begin{equation}
E_v(1) = -\mu \sum_i \sum_{m=-3}^3 |c_{i,(n=1,S=3,m)}|^2,
\end{equation}
while for $n=2$ the energy is given by
\begin{eqnarray}
E_v(2) &=& \sum_{i}~ \{ (U_0-2~\mu)~\sum_{S} \sum_{m=-S}^S |c_{i,(n=2,S,m)}|^2 \nonumber \\
&&+ \frac{U_2}{2} \sum_S \sum_{m=-S}^S |c_{i,(n=2,S,m)}|^2 (S(S+1)-24) \nonumber \\
&&+ U_1~|c_{i,(n=2,S=0,m=0)}|^2 \nonumber \\
&&+ \frac{U_3}{2} ( 288~|c_{i,(n=2,S=0,m=0)}|^2 \nonumber \\
&&~~~~~~+ 162 \sum_{m=-2}^2 |c_{i,(n=2,S=2,m)}|^2 \nonumber \\
&&~~~~~~+ 8 \sum_{m=-4}^4 |c_{i,(n=2,S=4,m)}|^2 \nonumber \\
&&~~~~~~+ 162 \sum_{m=-6}^6 |c_{i,(n=2,S=6,m)}|^2 ) \}.
\end{eqnarray}

The Mott phases are then obtained by minimizing $E_v(n)$ with respect to
the variational parameters $c_{i,(n,S,m)}$.

\end{document}